\documentclass[cfse]{cfse}

\pdfoutput=1

\usepackage{graphicx}
\usepackage{fancyvrb}
\usepackage{multirow}
\usepackage{hyperref}
\usepackage{times}

\toappear{1} 

\begin{document}


\title{JooFlux : modification de code à chaud et injection d'aspects directement dans une JVM 7}
\shorttitle{JooFlux}

\author{Julien Ponge et Frédéric Le Mouël}%

\address{Université de Lyon,\\
INSA-Lyon, CITI-INRIA F-69621, Villeurbanne, France\\
\texttt{julien.ponge@insa-lyon.fr} et \texttt{frederic.le-mouel@insa-lyon.fr}}

\date{\today}

\maketitle

\begin{abstract} 
  Modifier les portions fonctionnelles et non-fonctionnelles des applications au cours de leur
  exécution est utile et parfois critique tant lors des phases de développement que de suivi en
  production. JooFlux est un agent JVM qui permet à la fois de remplacer dynamiquement des
  implémentations de méthodes que d'appliquer des aspects. Il fonctionne en opérant des
  transformations de bytecode pour tirer parti de la nouvelle instruction \texttt{invokedynamic}
  apparue dans Java SE 7 pour faciliter l'implémentation de langages dynamiques au-dessus de la JVM.
  JooFlux peut être piloté à l'aide d'un agent JMX afin de conduire les modifications à l'exécution,
  et ce, sans s'appuyer sur un langage dédié. Nous avons comparé JooFlux à d'autres plates-formes d'AOP
  et des langages dynamiques. Les résultats montrent que les performances de JooFlux sont proches de
  celles du code Java non modifié. Le sur-coût est généralement négligeable, et nous observons même
  parfois de légers gains, tandis que les plates-formes d'AOP et langages dynamiques montrent des
  performances significativement dégradées. Ceci ouvre de prometteuses perspectives tant pour les
  évolutions futures de JooFlux que pour des études de cas.
    
  \MotsCles{JVM, patch à l'exécution, AOP, invokedynamic, bytecode}
\end{abstract}

\section{Introduction}

La  plupart  des   applications  sont  --  jusqu'à   un  certain  point  --   statiques.  Cela  est
particulièrement  vrai pour  les applications  développées en  utilisant des  langages statiquement
typés. Les langages typés dynamiquement offrent  souvent la possibilité de remplacement de méthodes
pendant  l'exécution, mais  ces changements  n'interviennent que  depuis le  programme lui-même  et
rarement d'outils  externes. Très peu  de langages offrent cet  accès extérieur, même  les langages
purement fonctionnels avec un shell interactif.

Les   modifications   à   chaud   de   code   ont   pourtant   de   nombreuses   et   intéressantes
vertus~\cite{Frieder1991,Hicks2005}. Pendant  les phases de  développement, un temps  important est
passé et  perdu à redémarrer les  applications pour observer  les effets des modifications  du code
source.  En  production,  l'application  de  correctifs  d'erreurs  ou  de  sécurité  requiert  des
redémarrages. Un redémarrage  complet peut prendre plusieurs minutes pour  être opérationnel quand,
dans  de nombreux  cas, une  simple correction  consiste juste  à remplacer  l'implémentation d'une
méthode sans introduire  d'effets de bord comme  un changement de signature. L'ajout  et le retrait
dynamiques de  fonctionnalités transversales, comme  des aspects de  sécurité ou de  trace, peuvent
également être  très utiles pour  capturer ou  surveiller ponctuellement certains  comportements de
l'application.

Dans le cas d'environnements adaptatifs d'exécution comme la machine virtuelle Java (JVM), ces
changements dynamiques impliquent malheureusement la perte des optimisations effectuées --
optimisations qui ne peuvent être obtenues et stabilisées qu'après une longue exécution --
conduisant donc à une dégradation des performances.

\paragraph{Notre proposition} Cet article décrit une nouvelle approche pour modifier "à chaud" des
méthodes et injecter des aspects pendant l'exécution d'applications Java. Nos contributions
s'articulent comme suit :

\begin{enumerate}

  \item Nous avons conçu et développé un agent Java qui intercepte et modifie le bytecode pendant
  l'exécution pour remplacer toutes les invocations de méthodes par le nouvel opcode
  \texttt{invokedynamic} introduit dans Java~7.
  
  \item Nous proposons une API de contrôle accessible par un agent~JMX qui permet de gérer ces
  modifications pendant l'exécution et ce pour pouvoir se dispenser d'un langage dédié ou d'insertion
  d'annotations ou de points de coupe dans le code source.

  \item Nous présentons une comparaison entre notre prototype JooFlux et plusieurs plates-formes à
  aspect et langages dynamiques existants. Les résultats mettent en exergue que les performances de
  JooFlux sont proches de celles de Java -- avec un coût marginal et même quelques fois un gain --
  tandis que les plates-formes à aspects et langages dynamiques présentent des sur-coûts importants.

\end{enumerate}

Cet article est structuré avec en premier le contexte AOP, JVM et le nouvel opcode
\texttt{invokedynamic}. Nous proposons ensuite JooFlux, son architecture et son fonctionnement avant
de donner une comparaison de ses performances. Après un bref état de l'art, nous concluons et
donnons des perspectives à ces travaux.

\section{Contexte}

Cette section comporte 3 parties. Nous commençons par quelques rappels informels sur la
programmation orientée-aspects. Nous faisons de même avec un aperçu de la machine virtuelle Java.
Enfin, nous nous focalisons sur une évolution de Java SE 7 qui est au c{\oe}ur de notre approche, et
dont l'origine est dans l'amélioration du support de langages dynamiques sur la JVM.

\subsection{Programmation orientée-aspects}

La programmation orientée-aspects (AOP) vise la modularisation des logiques applicatives
transversales, le plus souvent non fonctionnelles, telles la sécurité, la validation, les
transactions ou encore l'audit de traces. En effet, le code correspondant doit souvent être répété
sur différentes couches applicatives, ce qui augmente le coût de maintenance. Aussi, le code
fonctionnel se retrouve alors mêlé au code non fonctionnel.

L'AOP apporte une possible réponse à ses problèmes :
\begin{enumerate}
  \item un \emph{aspect} capture une logique applicative transversale en une unique unité de code,
    et
  \item un \emph{point de coupe} spécifie un ensemble de points de code exécutable, et
  \item un \emph{conseil} matérialise l'application de l'aspect au point de coupe.
\end{enumerate}

Supposons que l'on souhaite tracer chaque invocation d'un accesseur des classes du paquet
\texttt{foo}. Un aspect est alors le code de trace. Le point de coupe correspond aux appels des
méthodes publiques des classes de \texttt{foo} dont le nom commence par \texttt{get}. L'application
de l'aspect se fait avec un \emph{tisseur de code}, qui insère alors l'aspect aux points de coupe
concordants.

\subsection{La machine virtuelle Java}

La spécification de la machine virtuelle Java (JVM) est restée stable au cours des versions
successives de la plate-forme~\cite{LindholmJVM99}. Une JVM consomme du bytecode qui est
classiquement stocké dans des fichiers \texttt{.class} et qui correspondent chacun à une classe Java
compilée. Le format de ces classes compilées reflète la définition des classes dans le
langage source, à savoir que chaque \texttt{.class} contient une seule classe, ses attributs et ses
méthodes. On trouve également un \emph{constant pool} qui correspond à un ensemble indexé de valeurs
constantes qui peuvent être référencées depuis le reste du bytecode. Ceci permet de réduire
l'empreinte mémoire des classes en évitant des redondances, par exemple pour se référer à des chaînes
de caractères. Le modèle d'exécution de la JVM s'appuie sur une pile. Des \emph{opcodes} permettent
de la manipuler en consommant des opérandes puis poussant des résultats.

Même si le format du bytecode Java a été conçu de façon très proche de la structure du langage de
programmation éponyme, il demeure néanmoins un bytecode généraliste. Il est strictement plus
expressif que le langage source, ce qui a été exploité par des outils comme des
ofuscateurs~\cite{Batchelder2007}.
La JVM ne se limite pas à l'exécution d'applicatifs écrits en Java. Il existe un écosystème très
actif de langages s'exécutant sur la JVM. Ceci inclut tant le portage de langages déjà existants
(JRuby, Jython, Rhino) que de langages originaux (Groovy, Scala, Clojure). De nombreux travaux
d'optimisation des performances de la JVM couplés à une validation industrielle à large échelle font
de cette dernière une cible très intéressante pour des langages à la recherche d'un environnement
d'exécution performant~\cite{Paleczny2001,Kotzmann2008,Haubl2011}.

\subsection{Java SE 7 et \texttt{invokedynamic}}

La spécification de la JVM offrait 4 opcodes pour invoquer des méthodes jusqu'à la version 7.
\texttt{invokestatic} est utilisé pour les méthodes statiques. \texttt{invokevirtual} est utilisé
pour effectuer un appel en fonction du type de l'objet receveur, ce qui correspond aux méthodes
publiques et protégées et qui peuvent être surchargées dans des sous-classes. \texttt{invokespecial}
sert aux appels directs où le type est spécifié dans la signature, soit les constructeurs et
méthodes privées. Enfin, \texttt{invokeinterface} fait un appel par rapport à une définition
d'interface. Ce jeu d'opcodes est resté stable jusqu'à Java SE 7 qui a introduit
\texttt{invokedynamic}~\cite{Rose2009}.

\paragraph{Un nouvel opcode}
L'objectif principal de ce nouvel opcode d'invocation de méthode est de faciliter l'implémentation
de langages dynamiques sur la JVM. En effet, les implémentations de ces langages résolvent souvent
les types, symboles et cibles d'appel à l'exécution. Avec les opcodes antérieurs, les implémenteurs
de langages dynamiques devaient se tourner vers de la réflexivité et des proxys dynamiques pour
déferrer ces actions à l'exécution. L'impact sur les performances est significatif, et les
optimisations adaptatives (JIT) des machines virtuelles avaient le plus souvent du mal à s'appliquer
sur ces constructions.

\texttt{invokedynamic} est très proche de \texttt{invokeinterface} et \texttt{invokevirtual} dans le
sens où le choix de l'implémentation de méthode à exécuter se fait à l'exécution. Cependant ces 2
dernières font ce choix en fonction d'un receveur qui est l'objet dont la définition de classe
possède ladite méthode, soit en surcharge locale ou via la hiérarchie d'héritage.
\texttt{invokedynamic} relâche ces contraintes et se rapproche de pointeurs de fonctions comme en C.
Plus spécifiquement, un appel avec l'opcode \texttt{invokedynamic} se définit comme suit :
\begin{enumerate} 
  \item un nom symbolique pour désigner l'invocation, et
  \item une signature pour les paramètres et le type de retour, et
  \item une instruction d'initialisation qui est invoquée lors du premier passage sur le site
    d'appel de l'invocation.
\end{enumerate}

\paragraph{Une API pour la liaison dynamique de sites d'appels}
Le rôle de l'instruction d'initialisation est de lier le site d'appel avec une cible. Pour se faire, 
une nouvelle API a été définie dans le package \texttt{java.lang.invoke}. Elle fournit en particulier
2 types clé. \texttt{CallSite} représente un site d'appel. Il pointe vers une instance de
\texttt{MethodHandle} qui est soit une référence directe sur une méthode de classe ou attribut, soit
une chaîne de \emph{combinateurs}~\cite{Rose2009}. Voici un exemple d'utilisation de cette API :

{\footnotesize
\begin{Verbatim}[frame=single]
public static MethodHandle replaceSpaces(Lookup lookup) throws Throwable {
  return insertArguments(lookup.findVirtual(String.class, "replaceAll",
    methodType(String.class, String.class, String.class)), 1, "%20", " ");
}

public static void main(String... args) throws Throwable {
  MethodHandle mh = replaceSpaces(lookup());
  System.out.println((String) mh.invokeExact("A%20B%20C%20"));
}
\end{Verbatim}
}

Il  illustre  comment  obtenir  une référence  de  méthode  sur  \texttt{String.replaceAll(String,
String)}.  Ensuite, il  utilise  le combinateur  \texttt{insertArguments}  pour pré-lier  certains
arguments à  des valeurs constantes.  Ici, nous  lions les 2  arguments de sorte  que l'invocation
remplace les  occurrences de  \texttt{"\%20"} par  \texttt{" "}. Comme  la méthode  est virtuelle,
l'argument d'index 0  est le receveur. L'invocation  finale sur la chaîne d'exemple  donne bien un
affichage de \texttt{"A B  C"}. Un plus large panel de combinateurs est  proposé afin de faire des
transformations et  adaptations depuis une  cible d'appel vers  une cible effective.  Nous verrons
plus loin une mise en pratique pour mettre en oeuvre des fonctionnalités de JooFlux.

\begin{figure}[ht]
  \centering
  \includegraphics[width=0.65\textwidth]{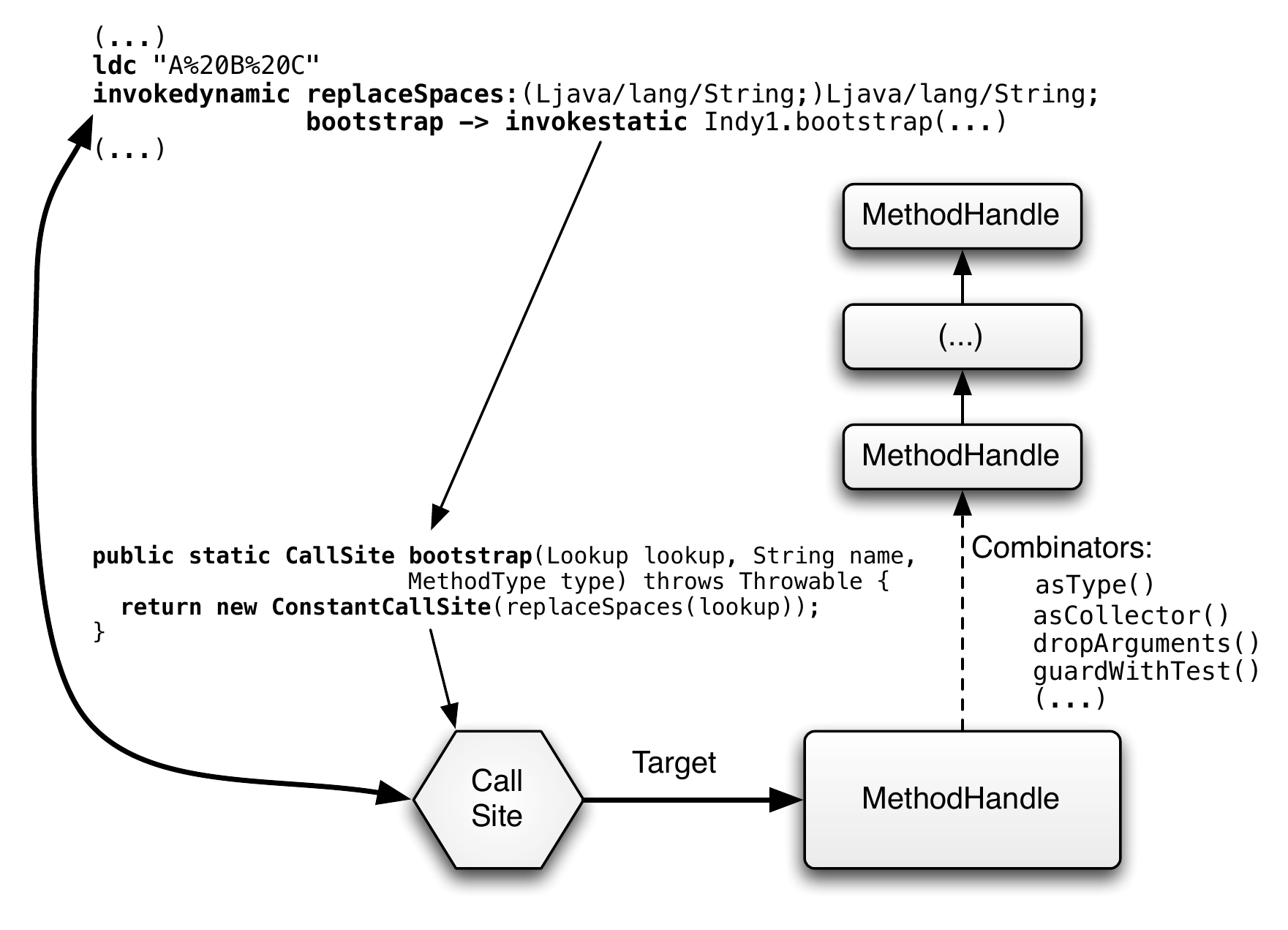}
  \caption{Initialisation d'un site d'appel \texttt{invokedynamic}.}
  \label{fig:indy-bootstrap}
\end{figure}

Le lien entre un site d'appel \texttt{invokedynamic} et l'appel d'initialisation est fréquemment
fait à l'aide d'une méthode statique qui construit une référence de méthode dont le type direct ou
transformé correspond au site d'appel, et enfin retourne une instance de \texttt{CallSite}. Ceci est
illustré par la Figure~\ref{fig:indy-bootstrap}. L'instance de \texttt{CallSite} reste constante
pour un site d'appel donné. En revanche, sa cible de référence de méthode peut changer.

Une des améliorations principales avec \texttt{invokedynamic} est que les chaînes de combinateurs sont
liées à des sites d'appels clairement identifiés et qui ont des branchements internes dans la JVM.
De la sorte, des optimisations se mettent en place plus efficacement qu'avec les approches
classiques basées sur la réflexivité~\cite{Thalinger2010}. Enfin, la vérification de type ne se fait
qu'à la création d'une référence de méthode et/ou d'un combinateur, à l'inverse de la réflexivité
classique où cela doit être fait pour chaque appel. Ceci aide à l'amélioration des performances.

\section{JooFlux}

Cette section comporte les détails techniques du fonctionnement de JooFlux. Nous expliquons comment
une indirection lors des appels de méthodes est introduite à l'aide de transformation de bytecode et
de \texttt{invokedynamic}. Nous expliquons ensuite comment les aspects peuvent être attachés aux
méthodes. Enfin nous présentons l'interface de gestion de JooFlux.

\subsection{Introduction d'une indirection}

JooFlux introduit une indirection sur les appels de méthode, de sorte que le remplacement de méthode
et l'application d'aspects puisse se faire à l'exécution. Si JooFlux cible principalement le
bytecode émis par un compilateur pour le langage Java, il peut théoriquement s'appliquer sur
n'importe quel bytecode valide produit par un langage autre que Java sur la JVM tel 
Scala~\cite{Odersky05}.
Le fait de pouvoir modifier les références de méthodes avec \texttt{invokedynamic} avec de nouvelles
chaînes de combinateurs permet de changer les implémentations de méthodes. Le jeu de combinateurs
disponibles nous permet également de greffer des traitements supplémentaires pré et post invocation
sur les arguments et valeurs de retour. Nous allons voir que ceci nous permet de faire de la
programmation par aspects.
Il est possible d'attacher des \emph{agents} à une JVM. Ils ont la possibilité d'effectuer diverses
opérations dont l'interception et transformation de bytecode au moment où il va être chargé dans la
JVM.

\begin{figure}[ht]
  \centering
  \includegraphics[width=\textwidth]{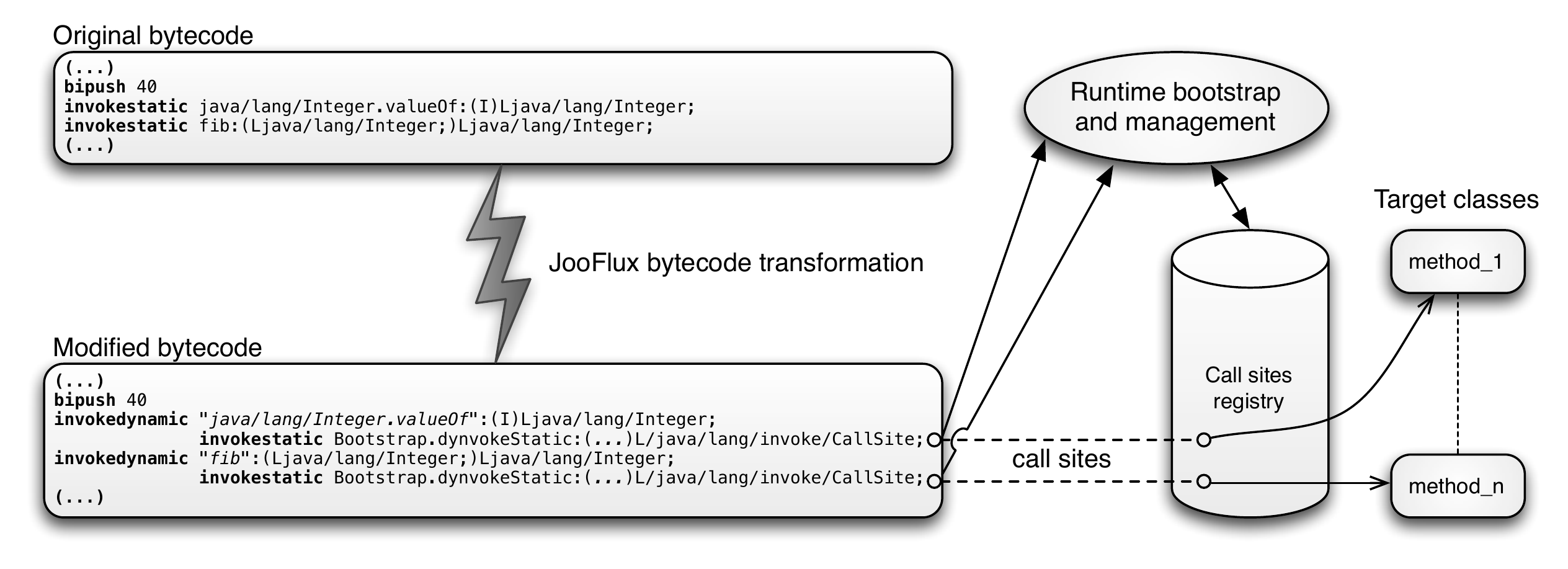}
  \caption{Aperçu de l'agent JooFlux pour la JVM.}
  \label{fig:jooflux-transfo}
\end{figure}

La Figure~\ref{fig:jooflux-transfo} donne un aperçu du fonctionnement de JooFlux en tant qu'agent
JVM. Il cherche les occurrences de \texttt{invokestatic}, \texttt{invokevirtual},
\texttt{invokespecial} ou \texttt{invokeinterface} puis les remplace par une construction
sémantiquement équivalente basée sur \texttt{invokedynamic}. Les sites d'appels sont liés à une
référence de méthode au moment de l'exécution. Les transformations de bytecode s'opèrent à l'aide de
la bibliothèque ASM\footnote{ASM 4.0: \url{http://asm.ow2.org/}}.
Initialement, les sites d'appels sont liés à des références de méthodes pointant sur les méthodes
des classes originelles. Ainsi pour un \texttt{invokestatic} dans le bytecode, un remplacement est
fait avec \texttt{invokedynamic} et un nom symbolique basé sur la signature de l'instruction
\texttt{invokestatic} remplacée. Ceci permet d'avoir un schéma de nommage uniforme. Le type de
l'invocation originelle est également préservée. Coté JooFlux, l'initialisation de sites d'appels se
fait via un jeu de méthodes statiques spécialisées en fonction du type d'invocation initiale. Ceci
introduit une couche d'indirection fine.

\subsection{Aspects via des combinateurs de références de méthodes}

Une fois les classes transformées par JooFlux et chargées dans la machine virtuelle, JooFlux
n'effectue plus de modifications de bytecode. L'injection se fait au niveau des chaines de
combinateurs référencées par les sites d'appels. JooFlux utilise 2~combinateurs que l'on trouve dans 
la classe \texttt{java.lang.invoke.MethodHandles} :
\begin{enumerate}
  \item \texttt{filterArguments} prend pour argument une référence de méthode cible, un index
    d'argument et un tableau de références de méthodes à utiliser en filtres, et
  \item \texttt{filterReturnValue} prend pour argument une référence de méthode cible et une
    référence de méthode filtre.
\end{enumerate}

Les types des références de méthodes filtres doivent correspondre à ceux des méthodes cibles. Dans
notre cas nous avons choisi une approche générique pour qu'un aspect soit agnostique de la signature
de la méthode interceptée. Dans le cas des appels de méthode les arguments sont capturés dans un
tableau d'objets, et dans le cas des retours un objet est employé. La classe d'exemple suivante peut
être utilisée pour obtenir des traces sur des appels de méthodes :

{\footnotesize
\begin{Verbatim}[frame=single]
public class Dumpers {
  public static Object[] onCall(Object[] args) {
    System.out.println(">>> " + Arrays.toString(args));
    return args;
  }
  public static Object onReturn(Object retval) {
    System.out.println("<<< " + retval);
    return retval;
  }
}
\end{Verbatim}
}

Plus généralement, les aspects peuvent faire d'autres choses comme lever une exception si un
argument ne respecte pas une pré-condition. Étant donné qu'ils agissent comme des filtres, ils
peuvent aussi modifier les arguments avant de les passer à la méthode cible ainsi que modifier la
valeur de retour.
Les aspects peuvent être empilés sur un même site d'appel. Par exemple, il est possible d'empiler
des aspects de validation et de trace. Ce qu'un aspect peut faire dépend aussi du type d'invocation
originelle. En effet, dans le cas d'une invocation statique les arguments sont ceux de la méthode.
En revanche pour les autres types d'invocation le premier argument est le receveur, autrement dit
l'objet sur lequel la méthode est invoquée.

Enfin, nous avions mentionné que les types des méthodes filtres doivent concorder avec ceux des
méthodes cibles. Dans notre cas nous devons effectuer des adaptations pour aller vers des types
génériques. Nous utilisons pour cela 2 combinateurs. Le premier est \texttt{asSpreader} qui collecte
des arguments dans un tableau. Le deuxième, \texttt{asCollector}, effectue l'opération inverse pour
distribuer les valeurs d'un tableau vers des paramètres. Dans notre cas, nous convertissons les
arguments d'une méthode vers un tableau de type \texttt{Object[]} avec \texttt{asSpreader}. Celui-ci
est passé au filtre, invoquant donc le conseil de l'aspect. Ensuite nous utilisons
\texttt{asCollector} pour réadapter vers la méthode cible originale. La gestion de l'interception
des retours de méthode est plus simple et se résume juste à passer d'un type spécifique à
\texttt{Object} pour le conseil, puis à raffiner le type vers celui d'origine. Ces adaptations de
type se font avec le combinateur \texttt{asType} et permettent de rendre cohérents les types sur une
chaîne de combinateurs.

\subsection{Gestion des sites d'appel}

Les sites d'appel sont mis dans un registre central lors de leur création. Le registre a pour seul
rôle de garder une référence sur les sites d'appel, et ce, afin de pouvoir ultérieurement les
modifier pour appliquer un aspect. En revanche, chaque site d'appel est bien initialisé lors de la
première invocation de son instruction \texttt{invokedynamic}, et les appels suivants sont
directement résolus via la chaine de combinateurs associée. Par conséquent, le registre n'intervient
pas lors des appels de méthode et ne possède pas de coût intrinsèque sur les performances des appels
de méthodes.

Il existe deux types de sites d'appels offerts par \texttt{java.lang.invoke} dont la cible puisse
changer en cours d'exécution : \texttt{MutableCallSite} et
\texttt{VolatileCallSite}. Leur sémantique correspond respectivement à un attribut de classe normal
et à un attribut de classe \texttt{volatile} tels que spécifiés dans la \emph{JSR 133} \footnote{JSR
133: JavaTM Memory Model and Thread Specification -- \url{http://jcp.org/en/jsr/detail?id=133}}.

JooFlux utilise des instances de \texttt{VolatileCallSite}. Ainsi, toute modification effectuée sur un
site d'appel est visible par tous les threads qui souhaiteraient effectuer une invocation. À
l'inverse, l'emploi de \texttt{MutableCallSite} est sujet à une mise en cache par un \emph{thread},
et une modification d'un site d'appel n'est alors pas nécessairement immédiatement visible des
autres \emph{threads}. Ce choix a un impact potentiel sur les performances, principalement en cas de
contention avec une écriture. Les résultats dépendent grandement de l'architecture matérielle
utilisée et des tentatives d'optimisation du JIT. 

\subsection{Piloter JooFlux pour remplacer à chaud des méthodes et injecter des aspects}

La couche de gestion de JooFlux utilse le registre de diverses façons via un agent
JMX\footnote{Java Management Extensions:
\url{http://jcp.org/aboutJava/communityprocess/final/jsr003/index3.html}} pour effectuer des
requêtes.

Cet agent offre les opérations distantes suivantes :
\begin{enumerate}
  \item remplacer les implémentations de méthode, et
  \item injecter un aspect avant ou après certains sites d'appel, et
  \item obtenir diverses métriques telles le nombre de sites d'appels gérés par JooFlux ou leur
    liste.
\end{enumerate}

À titre d'exemple, voyons l'interaction pour un remplacement de méthode via l'interface de l'agent
JMX :

{\footnotesize
\begin{Verbatim}[frame=single]
void changeCallSiteTarget(String methodType, 
                          String oldTarget, String newTarget)
\end{Verbatim}
}

Le paramètre \texttt{methodType} spécifie le type d'invocation et prend pour valeur static, virtual
interface ou special. Les 2 autres paramètres spécifient un identifiant de site d'appel à remplacer,
et une référence de méthode pour la nouvelle cible.

\begin{figure}[!htb]
  \centering
  \includegraphics[width=0.8\textwidth]{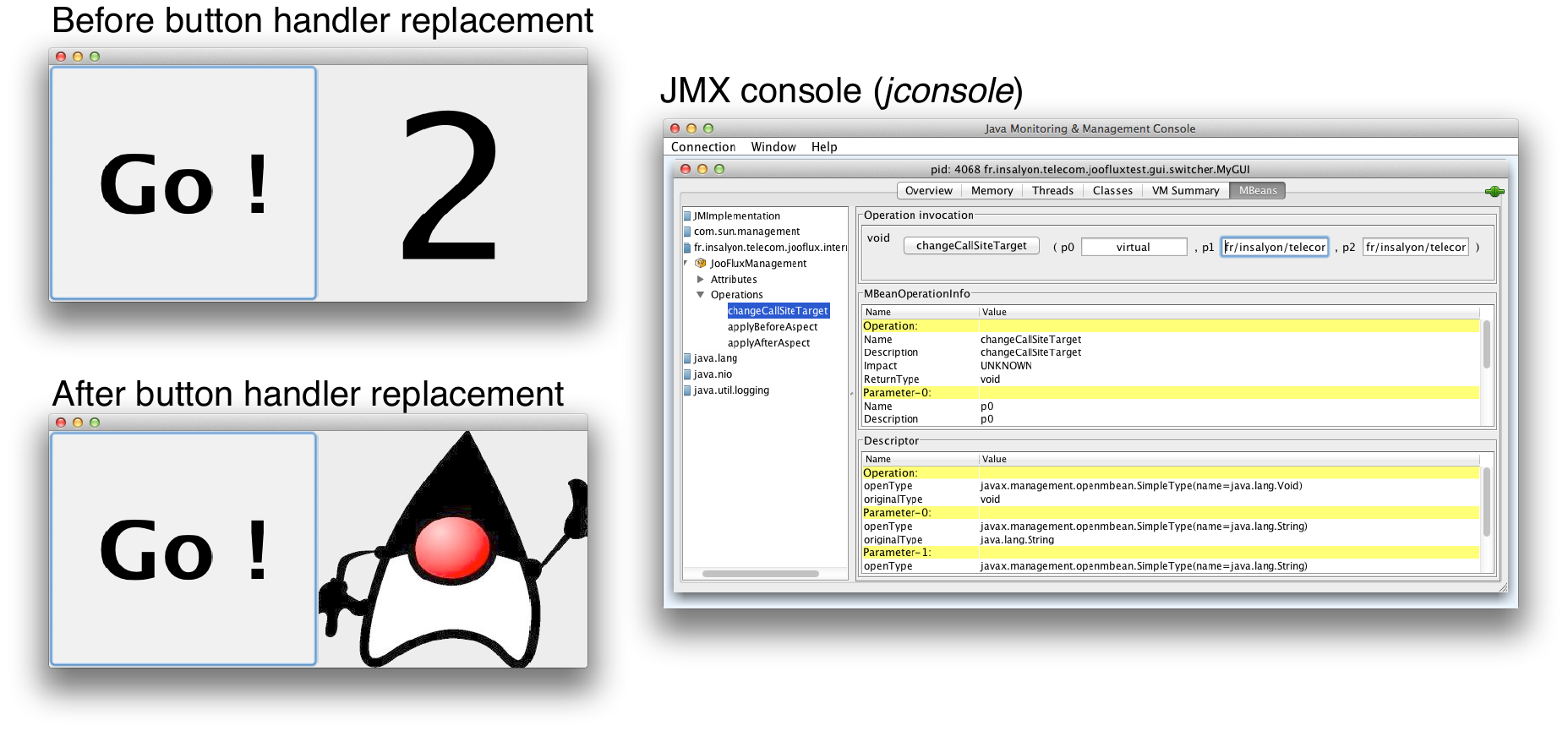}
  \caption{Interaction via JMX avec une application modifiée par JooFlux.}
  \label{fig:mbean}
\end{figure}

La Figure~\ref{fig:mbean} expose une application avec un bouton à gauche et un texte à droite. En
se connectant à l'agent JMX JooFlux via un outil comme \texttt{jconsole}, nous pouvons remplacer le
gestionnaire d'évènement de bouton de sorte qu'au prochain clic, une image remplace le texte. Dans
ce cas précis, nous avons utilisé les paramètres suivants pour l'opération
\texttt{changeCallSiteTarget} de l'agent JMX :

{\footnotesize
\begin{Verbatim}[frame=single]
virtual, // virtual method
fr/insalyon/telecom/joofluxtest/gui/switcher/
   MyActionListener.counterIncrement:(MyActionListener)void, // handler id
fr/insalyon/telecom/joofluxtest/gui/switcher/
   MyActionListener.pictureSwitch:()V // our new handler
\end{Verbatim}
}

Enfin, injecter un aspect via l'agent JMX se fait de façon similaire :

{\footnotesize
\begin{Verbatim}[frame=single]
void applyBeforeAspect(String callSitesKey,
                       String aspectClass, String aspectMethod)

void applyAfterAspect(String callSitesKey, 
                      String aspectClass, String aspectMethod)
\end{Verbatim}
}


\section{Comparaison expérimentale avec des plates-formes AOP et des langages dynamiques}
\label{section:bench}

Pour valider notre approche, nous comparons les performances de la machine virtuelle Java avec notre agent JooFlux à d'autres approches effectuant également de la re-direction dynamique de méthodes: les plates-formes à aspects~(section~\ref{section:aop}) et les langages de programmation dynamiques~(sections~\ref{section:microdynlang}-\ref{section:macrodynlang}). 

Toutes ces approches sont basées sur la machine virtuelle et ont été testées avec des micro- et des macro-jeux de tests sur un MacBook~Pro~2,3~GHz~Intel~Core2~(i5), 4~Go~1333~MHz~DDR3, avec Mac~OS~X~Lion~10.7.4~(11E53) et l'OpenJDK Runtime Environment~1.7.0-u10-b06-20120906\footnote{OpenJDK Runtime Environment~1.7.0-u10-b06-20120906: \url{http://code.google.com/p/openjdk-osx-build/}}. Ces tests ont été renouvelés 10~fois pour constituer le jeu de résultats, sur lequel nous avons calculé les quartiles et les sur-coûts médians.

\subsection{Micro-tests et plates-formes AOP}
\label{section:aop}

Nous comparons la fonctionnalité d'injection d'aspects de JooFlux avec deux plates-formes AOP : \mbox{AspectJ}\footnote{AspectJ~1.7.1: \url{http://www.eclipse.org/aspectj/}}\cite{KiczalesHHKPG01} et \mbox{Byteman}\footnote{Byteman~2.1.0: \url{http://www.jboss.org/byteman/}}\cite{Dinn2011}. Nous utilisons le calcul classique d'une fonction Fibonacci récursive (\texttt{classicfibo})\footnote{Recursive Fibonacci: \url{http://en.wikipedia.org/wiki/Recursion\_(computer\_science)\#Fibonacci}} à laquelle nous injectons avant et/ou après un aspect vide -- c.à.d. une re-direction d'appel vers une méthode vide. Le tableau~\ref{tab:fiboaop} présente les résultats.

\begin{table}[!htb]
\centering
{\footnotesize
\begin{tabular}{|c|c|c|c|c|c|c|c||c||}
\hline
\hline
Plat. Exec. & Tissage & Impl. & Q1\footnotesize{-min} & Q2\footnotesize{-25\%} & Q3\footnotesize{-median} & Q4\footnotesize{-75\%} & Q5\footnotesize{-max} & Sur-coût \\
\hline
\hline
JVM & - & classicfibo & 611 & 613 & 615 & 616 & 636 & - \\
\hline
& \multirow{7}{*}{compilation} & classicfibo & \multirow{2}{*}{609} & \multirow{2}{*}{613} & \multirow{2}{*}{620} & \multirow{2}{*}{621} & \multirow{2}{*}{650} & \multirow{2}{*}{+0,8\%} \\ 
& & + aspect avant & & & & & & \\
\cline{3-9}
\multirow{2}{*}{JVM +} & & classicfibo & \multirow{2}{*}{609} & \multirow{2}{*}{611} & \multirow{2}{*}{620} & \multirow{2}{*}{621} & \multirow{2}{*}{693} & \multirow{2}{*}{+0,8\%} \\
\multirow{2}{*}{AspectJ} & & + aspect après & & & & & & \\
\cline{3-9}
& & classicfibo & \multirow{3}{*}{609} & \multirow{3}{*}{613} & \multirow{3}{*}{620} & \multirow{3}{*}{622} & \multirow{3}{*}{717} & \multirow{3}{*}{+0,8\%} \\
& & + aspect avant & & & & & & \\
& & \& aspect après & & & & & & \\
\hline
& \multirow{7}{*}{exécution} & classicfibo & \multirow{2}{*}{3007} & \multirow{2}{*}{3028} & \multirow{2}{*}{3051} & \multirow{2}{*}{3070} & \multirow{2}{*}{3525} & \multirow{2}{*}{+396\%} \\ 
& & + aspect avant & & & & & & \\
\cline{3-9}
\multirow{2}{*}{JVM +} & & classicfibo & \multirow{2}{*}{1269} & \multirow{2}{*}{1273} & \multirow{2}{*}{1280} & \multirow{2}{*}{1295} & \multirow{2}{*}{1356} & \multirow{2}{*}{+108\%} \\
\multirow{2}{*}{Agent JooFlux} & & + aspect après & & & & & & \\
\cline{3-9}
& & classicfibo & \multirow{3}{*}{3804} & \multirow{3}{*}{3830} & \multirow{3}{*}{3861} & \multirow{3}{*}{3893} & \multirow{3}{*}{3950} & \multirow{3}{*}{+528\%} \\
& & + aspect avant & & & & & & \\
& & \& aspect après & & & & & & \\
\hline
& \multirow{7}{*}{exécution} & classicfibo & \multirow{2}{*}{79309} & \multirow{2}{*}{84998} & \multirow{2}{*}{85371} & \multirow{2}{*}{85695} & \multirow{2}{*}{86397} & \multirow{2}{*}{+13781\%} \\ 
& & + aspect avant & & & & & & \\
\cline{3-9}
\multirow{2}{*}{JVM +} & & classicfibo & \multirow{2}{*}{82561} & \multirow{2}{*}{82780} & \multirow{2}{*}{82911} & \multirow{2}{*}{83040} & \multirow{2}{*}{83406} & \multirow{2}{*}{+13381\%} \\
\multirow{2}{*}{Byteman} & & + aspect après & & & & & & \\
\cline{3-9}
& & classicfibo & \multirow{3}{*}{179984} & \multirow{3}{*}{188088} & \multirow{3}{*}{188187} & \multirow{3}{*}{188321} & \multirow{3}{*}{188971} & \multirow{3}{*}{+30499\%} \\
& & + aspect avant & & & & & & \\
& & \& aspect après & & & & & & \\
\hline
\hline
\end{tabular}
}
\caption{Fibo(40) micro-test pour les plates-formes AOP (en ms)}
\label{tab:fiboaop}
\end{table}

Le temps d'exécution du bytecode généré par \mbox{AspectJ} est proche de celui de Java, seulement il
tisse les aspects statiquement pendant la compilation et plus aucune modification ne peut être
effectuée à l'exécution. \mbox{Byteman} permet l'injection d'aspects pendant l'exécution en
déchargeant la classe et modifiant son bytecode pendant son rechargement. Cette technique dégrade
significativement les performances et implique que toutes les optimisations JIT sont perdues. Le
bytecode généré par \mbox{JooFlux} contient déjà les appels dynamiques, donc le coût de tissage d'un
aspect est seulement le coût d'ajout d'un nouveau combinateur dans la chaîne d'un site d'appel, de
recopier les arguments et de transférer la valeur de retour. En conservant la chaîne d'appel à peu
près intacte, \mbox{JooFlux} sauvegarde les optimisations JIT.

\subsection{Micro-tests et langages dynamiques}
\label{section:microdynlang}

De nombreux langages de programmation proposent la re-direction dynamique de méthodes directement dans le langage, par exemple avec les APIs de réflexion. Pour tester la fonctionnalité de re-direction dynamique de méthodes de JooFlux, nous la comparons aux principaux langages dynamiques basés sur la machine virtuelle : Java\footnote{Java~7: \url{http://docs.oracle.com/javase/7/docs/api/}}, Clojure\footnote{Clojure~1.4.0: \url{http://clojure.org}}, JRuby\footnote{JRuby~1.6.7.2~(-indy), 1.7.0.preview2~(+indy): \url{http://jruby.org}}, Groovy\footnote{Groovy~2.0.2~(-+indy): \url{http://groovy.codehaus.org}}, Rhino~JavaScript\footnote{Rhino~Javascript~1.7R4: \url{https://developer.mozilla.org/en-US/docs/Rhino}} and Jython\footnote{Jython~2.5.3: \url{http://www.jython.org}}. 

Nous testons la performance avec le même micro-jeu de test Fibonacci. Selon le langages de programmation, la fonction Fibonacci peut s'écrire de différentes manières : \texttt{classicfibo} manipule des objets, \texttt{fastfibo} manipule des paramètres de type long, \texttt{fastestfibo} manipule également des paramètres et une valeur résultat de type long, et finalement \texttt{reflectivefibo} utilise l'API de réflexion pour l'invocation de méthode. Ces différentes implémentations ont leur importance car elles influencent le bytecode généré et donc les performances.
Les résultats pour Java sont présentés dans le tableau~\ref{tab:fibojava}. Le sur-coût de JooFlux est insignifiant pour \texttt{classicfibo} et présente un facteur de ralentissement de~2 pour \texttt{reflectivefibo}. Cette mise en {\oe}uvre de Fibonacci est en effet notre pire cas, car la machine virtuelle peut difficilement effectuer des inclusions de code~(\emph{inlining}).


\begin{table}[!htb]
\centering
{\footnotesize
\begin{tabular}{|c|c|c|c|c|c|c|c||c||}
\hline
\hline
Lang. Prog. & Plat. Exec. & Impl. & Q1\footnotesize{-min} & Q2\footnotesize{-25\%} & Q3\footnotesize{-median} & Q4\footnotesize{-75\%} & Q5\footnotesize{-max} & Sur-coût \\
\hline
\hline
\multirow{2}{*}{Java} & \multirow{2}{*}{JVM} & classicfibo & 611 & 613 & 615 & 616 & 636 & - \\
\cline{3-9}
& & reflectivefibo & 1758 & 1762 & 1782 & 1803 & 4121 & - \\
\hline
\multirow{2}{*}{Java} & JVM + & classicfibo & 611 & 613 & 616 & 618 & 690 & +0,001\% \\ 
\cline{3-9}
& Agent JooFlux & reflectivefibo & 3668 & 3686 & 3717 & 3743 & 4273 & +108\% \\
\hline
\hline	
\end{tabular}
}
\caption{Fibo(40) micro-test pour le langage Java (en ms)}
\label{tab:fibojava}
\end{table}

Les résultats pour les autres langages de programmation sont présentés dans le
tableau~\ref{tab:fibodynlang}. Les sur-coûts pour \texttt{classicfibo}, \texttt{fastfibo} et \texttt{fastestfibo} sont comparés à ceux de JVM+JooFlux \texttt{classicfibo}. Les implantations de \texttt{reflectivefibo} sont toujours comparées à celle de JVM+JooFlux \texttt{reflectivefibo}.

Même si l'implémentation fortement typée en Clojure présente seulement un facteur de ralentissement de 1.2-1.4, la plupart des autres langages sont de 3~à~18~fois plus lents que notre prototype JooFlux. JRuby et Groovy proposent des versions bêta utilisant l'opcode \texttt{invokedynamic} \textit{(+indy, chiffres en italiques dans le tableau~\ref{tab:fibodynlang})} mais même s'ils gagnent un facteur 1.5-2, ils restent significativement plus lents que Java+JooFlux.

\begin{table}[!htb]
\centering
{\footnotesize
\begin{tabular}{|c|c|c|c|c|c|c|c||c||}
\hline
\hline
Lang. Prog. & Plat. Exec. & Impl. & Q1\footnotesize{-min} & Q2\footnotesize{-25\%} & Q3\footnotesize{-median} & Q4\footnotesize{-75\%} & Q5\footnotesize{-max} & JooFlux Diff \\
\hline
\hline
\multirow{3}{*}{Clojure} & \multirow{3}{*}{JVM} & fastestfibo & 722 & 732 & 734 & 740 & 742 & +19\% \\
\cline{3-9}
& & fastfibo & 859 & 862 & 864 & 875 & 892 & +40\% \\
\cline{3-9}
& & classicfibo & 4105 & 4118 & 4171 & 4265 & 4326 & +577\% \\
\hline
\multirow{4}{*}{JRuby} & \multirow{3}{*}{JVM} & \multirow{2}{*}{classicfibo} & 6290 & 6333 & 6382 & 6486 & 7043 & +936\% \\
& & & \it{(3982)} & \it{(4006)} & \it{(4020)} & \it{(4069)} & \it{(4323)} & \it{(+552\%)} \\
\cline{3-9}
& \it{(+indy)} & \multirow{2}{*}{reflectivefibo} & 10226 & 12020 & 12060 & 12076 & 12288 & +224\% \\
& & & \it{(7545)} & \it{(7561)} & \it{(7581)} & \it{(7621)} & \it{(7737)} & \it{(+104\%)} \\
\hline
\multirow{8}{*}{Groovy} & \multirow{8}{*}{JVM} & \multirow{2}{*}{fastestfibo} & 1383 & 1388 & 1394 & 1401 & 1417 & +126\% \\
& & & \it{(3061)} & \it{(3077)} & \it{(3092)} & \it{(3150)} & \it{(3165)} & \it{(+402\%)} \\
\cline{3-9}
& & \multirow{2}{*}{fastfibo} & 2709 & 2721 & 2725 & 2749 & 2766 & +342\% \\
& & & \it{(2513)} & \it{(2519)} & \it{(2528)} & \it{(2540)} & \it{(2583)} & \it{(+310\%)} \\
\cline{3-9}
& \multirow{2}{*}{\it{(+indy)}} & \multirow{2}{*}{classicfibo} & 8660 & 8691 & 8716 & 8726 & 9066 & +1315\% \\
& & & \it{(4461)} & \it{(4488)} & \it{(4522)} & \it{(4584)} & \it{(4656)} & \it{(+634\%)} \\
\cline{3-9}
& & \multirow{2}{*}{reflexivefibo} & 57734 & 57892 & 58009 & 58182 & 58364 & +1460\% \\
& & & \it{(8366} & \it{(8378)} & \it{(8386)} & \it{(8405)} & \it{(8697)} & \it{(+125\%)} \\
\hline
Javascript & JVM & classicfibo & 9052 & 9208 & 11275 & 11441 & 11764 & +1730\% \\
\hline
Jython & JVM & classicfibo & 29053 & 29258 & 29675 & 30202 & 31871 & +4717\% \\
\hline
\hline	
\end{tabular}
}
\caption{Fibo(40) micro-tests pour les langages de programmation basés sur la JVM (en ms)}
\label{tab:fibodynlang}
\end{table}

Pour les micro-tests, que ce soit AOP ou langages dynamiques, notre agent JooFlux réécrit le bytecode en 75-100ms pour une transformation de 2~classes~(\texttt{Fibonacci} et \texttt{InvokeBootstrap} -- la classe de
bootstrap est elle-même dynamique) et 1~méthode~(\texttt{classicfibo}). Ce délai raisonnable est introduit avant le démarrage de l'application et reste comparable aux délais de lancement des autres plates-formes.

\subsection{Macro-tests de calculs et réécritures intensifs}
\label{section:macrodynlang}

Pour tester plus intensément la redirection dynamique de méthodes avec JooFlux, nous avons utilisé 3~macro-jeux de tests~:
\begin{enumerate}
\item SCImark 2.0\footnote{SCImark 2.0: \url{http://math.nist.gov/scimark2/}}: SciMark 2.0 est un jeu de tests en Java pour le calcul scientifique et numérique, assez consommateur en CPU et mémoire. Il effectue plusieurs types de calcul et retourne un score composite en Mflops.
\item Un compteur de mots pouvant s'exécuter en parallèle avec Fork/Join~\cite{ponge:inria-00611456}: ce compteur stresse particulièrement la mémoire et les entrées/sorties en comptant le nombre de mots dans chaque fichier contenu dans une arborescence de fichiers.
\item L'environnement d'exécution du langage Clojure au-dessus de JooFlux: nous avons utilisé le langage de programmation Clojure dont l'implantation est entièrement écrite en Java pour avoir un jeu de tests effectuant un nombre massif de réécritures de bytecode.
\end{enumerate}

\paragraph{SCImark 2.0}

Ce jeu de tests effectue des transformations de Fourier (\emph{Fast Fourier Transformations}, FFT), résout des équations de Laplace (\emph{Jacobi Successive Over-relaxation}, SOR), approxime la valeur de $\pi$ (\emph{Monte Carlo integration}, MC), multiplie des matrices (\emph{Sparse matrix multiply}, SM), effectue des factorisations de matrices denses (\emph{dense LU matrix factorization}, LU) et calcule finalement un score composé~(\emph{Composite Score}, CS). Ces différents calculs sont réalisés sur un petit ou un grand jeu de données.

Les performances avec et sans JooFlux sont présentées dans le tableau~\ref{tab:scimark2java}. La plupart des calculs présentent des sur-coûts ou gains marginaux, excepté pour l'intégration Monte Carlo qui perd 20\% en performance. L'algorithme Monte Carlo utilise des fonctions synchronisées et de l'inclusion de fonctions (cf documentation de SCImark) et les fonctions synchronisés n'impactant pas les performances (cf le macro-test du comptage parallèle ci-après), la perte provient de difficultés d'optimisations JIT du fait de l'inclusion de fonctions.

\begin{table}[!htb]
\centering
{\footnotesize
\begin{tabular}{|c|c|c|c|c|c|c|c|c|c|c|c|}
\hline
\hline
& \multicolumn{2}{c|}{FFT} & \multicolumn{2}{c|}{SOR} & MC & \multicolumn{2}{c|}{SM} & \multicolumn{2}{c|}{LU} & \multicolumn{2}{c|}{CS} \\
\cline{2-12}
& \multirow{2}{*}{$\!\!\!2^{10}\!\!\!$} & \multirow{2}{*}{$\!\!\!2^{20}\!\!\!$} & \multirow{2}{*}{$\!\!10^{2}\!\!\times\!\!10^{2}\!\!$} & \multirow{2}{*}{$\!\!10^{3}\!\!\times\!\!10^{3}\!\!$} & \multirow{2}{*}{-} & $\!\!\!N\!\!=\!\!10^{3}\!\!\!$ & $\!\!\!N\!\!=\!\!10^{5}\!\!\!$ & \multirow{2}{*}{$\!\!10^{2}\!\!\times\!\!10^{2}\!\!$} & \multirow{2}{*}{$\!\!10^{3}\!\!\times\!\!10^{3}\!\!$} & \multirow{2}{*}{Small} & \multirow{2}{*}{Large} \\
& & & & & & $\!\!nz\!\!=\!\!5.10^{3}\!\!\!$ & $\!\!\!nz\!\!=\!\!10^{6}\!\!\!$ & & & & \\
\hline
JVM & 827 & 178,7 & 1196,5 & 1077,6 & 635,4 & 1180,8 & 1227,8 & 2467,8 & 1490,2 & 1261,5 & 915,4 \\
\hline
JVM + & \multirow{2}{*}{824,3} & \multirow{2}{*}{178,5} & \multirow{2}{*}{1193,2} & \multirow{2}{*}{1080,3} & \multirow{2}{*}{507,2} & \multirow{2}{*}{1172,9} & \multirow{2}{*}{1249,5} & \multirow{2}{*}{2325,9} & \multirow{2}{*}{1489,7} & \multirow{2}{*}{1204,9} & \multirow{2}{*}{900,2} \\
JooFlux & & & & & & & & & & & \\
\hline
\hline	
Perf. & -0,3\% & -0,1\% & -0,3\% & +0,3\% & -20\% & -0,7\% & +1,8\% & -5,8\% & -0,03\% & -4,5\% & -1,7\% \\
\hline
\hline
\end{tabular}
}
\caption{SCImark2.0 macro-test (in Mflops)}
\label{tab:scimark2java}
\end{table}

\paragraph{Compteur parallèle de mots dans les fichiers}

Nous avons appliqué le compteur de mots sur l'arborescence de fichiers source de la machine virtuelle \mbox{HotSpot}~($2.10^{3}$~files) avec un processus mono-thread et avec un processus à 2-threads sur 2~c{\oe}urs du CPU. Comme le montre le tableau~\ref{tab:forkjoinjava}, de manière surprenante, le bytecode modifié par JooFlux améliore les performances. Ce résultat montre bien que les appels synchronisés de méthodes ne sont pas affectés par l'indirection \texttt{invokedynamic} et que le JIT peut même en bénéficier et parfaitement trouver de meilleures inclusions de code.

\begin{table}[!htb]
\centering
{\footnotesize
\begin{tabular}{|c|c|c|c|c|c|c||c||}
\hline
\hline
Plat. Exec. & Impl. & Q1\footnotesize{-min} & Q2\footnotesize{-25\%} & Q3\footnotesize{-median} & Q4\footnotesize{-75\%} & Q5\footnotesize{-max} & Sur-coût \\
\hline
\hline
\multirow{4}{*}{JVM} & wordcounter & \multirow{2}{*}{3707} & \multirow{2}{*}{3727} & \multirow{2}{*}{3733} & \multirow{2}{*}{3752} & \multirow{2}{*}{4769} & \multirow{2}{*}{-} \\
& 1 thread & & & & & & \\
\cline{2-8}
& wordcounter & \multirow{2}{*}{1826} & \multirow{2}{*}{1934} & \multirow{2}{*}{2161} & \multirow{2}{*}{2235} & \multirow{2}{*}{5949} & \multirow{2}{*}{-} \\
& fork/join 2 threads & & & & & & \\
\hline
& wordcounter & \multirow{2}{*}{3634} & \multirow{2}{*}{3646} & \multirow{2}{*}{3658} & \multirow{2}{*}{3689} & \multirow{2}{*}{4829} & \multirow{2}{*}{-2\%} \\ 
JVM + & 1 thread & & & & & & \\
\cline{2-8}
Agent JooFlux & wordcounter & \multirow{2}{*}{1825} & \multirow{2}{*}{1916} & \multirow{2}{*}{2029} & \multirow{2}{*}{2070} & \multirow{2}{*}{2407} & \multirow{2}{*}{-6\%} \\
& fork/join 2 threads & & & & & & \\
\hline
\hline	
\end{tabular}
}
\caption{Macro-test du compteur parallèle de mots pour Java (en ms)}
\label{tab:forkjoinjava}
\end{table}

\paragraph{Clojure au-dessus de JooFlux}

L'environnement d'exécution du langage Clojure est une application Java assez conséquente~($3.10^{3}$~classes). Nous l'avons utilisé comme jeu de tests pour appliquer des réécritures et interceptions massives. Nous avons ré-exécuté le micro-test Fibonacci, mais cette fois avec une totale réécriture du bytecode de Clojure : 1325~classes transformées, 26866~méthodes transformées et 19646~interceptions initiales de méthodes. Nous ne traçons pas toutes les interceptions de méthodes -- seulement la première lors du passage par le registre -- car les outils de traces en continu sont bien connus pour ralentir dramatiquement les performances.
Le tableau~\ref{tab:fiboclojure}\footnote{Fibo(40) en Clojure n'a pas exactement les mêmes
performances dans le tableau~\ref{tab:fiboclojure} et le tableau~\ref{tab:fibodynlang}. Pour
permettre une bonne comparaison, nous avons préféré présenter les résultats de Clojure+JVM et ceux
de Clojure+JVM+JooFlux issus du même test.} présente les résultats. Le sur-coût introduit est
toujours insignifiant par rapport au temps d'exécution total. Nous pouvons donc dire que rendre le
langage Clojure dynamique est intéressant\footnote{"Why Clojure doesn’t need invokedynamic, but it
might be nice" : \url{http://blog.fogus.me/2011/10/14/why-clojure-doesnt-need-invokedynamic-but-it-might-be-nice/}} avec un coût réduit à l'exécution. Un surcoût de 4s est par contre introduit pour la transformation du bytecode. Celui-ci n'est plus négligeable mais il s'applique seulement une fois au chargement, ne change en rien les performances pendant l'exécution et est un prix raisonnable pour rendre l'application entièrement dynamique.

\begin{table}[!htb]
\centering
{\footnotesize
\begin{tabular}{|c|c|c|c|c|c|c|c||c||}
\hline
\hline
Lang. Prog. & Plat. Exec. & Impl. & Q1\footnotesize{-min} & Q2\footnotesize{-25\%} & Q3\footnotesize{-median} & Q4\footnotesize{-75\%} & Q5\footnotesize{-max} & Sur-coût \\
\hline
\hline
\multirow{3}{*}{Clojure} & \multirow{3}{*}{JVM} & fastestfibo & 701 & 702 & 704 & 708 & 717 & - \\
\cline{3-9}
& & fastfibo & 839 & 845 & 848 & 851 & 855 & - \\
\cline{3-9}
& & classicfibo & 3969 & 3978 & 3984 & 4010 & 4093 & - \\
\hline
\multirow{3}{*}{Clojure} & \multirow{2}{*}{JVM +} & fastestfibo & 695 & 699 & 702 & 704 & 713 & -0,3\% \\
\cline{3-9}
& \multirow{2}{*}{Agent JooFlux} & fastfibo & 856 & 861 & 864 & 868 & 891 & +1,9\% \\
\cline{3-9}
& & classicfibo & 3957 & 3969 & 3980 & 4002 & 4069 & -0,1\% \\
\hline
\hline	
\end{tabular}
}
\caption{Fibo(40) en Clojure au dessus de la plate-forme JVM+JooFlux (en ms)}
\label{tab:fiboclojure}
\end{table}

\section{État de l'art}

\paragraph{Programmation par aspects} De nombreux travaux s'attaquent à la conception et
l'implantation de la programmation par aspects.  AspectJ est une plate-forme bien connue qui
comprend un langage d'aspect et un compilateur~\cite{KiczalesHHKPG01}. Bien que statique, elle
permet des transformations et tissages de bytecode efficaces, et offre un langage à granularité fine pour
définir les points de coupe. Javassist est une autre plate-forme AOP statique en
Java~\cite{Chiba2003}. Certains travaux se concentrent sur le tissage dynamique
d'aspects~\cite{Popovici2002,Popovici2003}. JBoss Byteman est particulièrement intéressant car il
propose des règles de type ECA~(Évènement-Condition-Action)~\cite{Dinn2011}. Par le biais d'un agent
Java qui peut être appliqué à un programme en cours d'exécution, il peut appliquer ou retirer ces
règles dynamiquement. Les travaux de~\cite{Bockisch2004} illustrent le fait que fournir une machine
virtuelle modifiée est une solution permettant d'offrir un panorama complet d'opérations de
modifications dynamiques d'applications Java. Ces résultats concordent avec~\cite{Wurthinger2011}.

\paragraph{Machines virtuelles} Le domaine des machines virtuelles ne manque pas de problématiques.
La JVM est une plate-forme attractive pour les travaux de recherche étant donné sa spécification
ouverte~\cite{LindholmJVM99}. De nombreux travaux visent à en améliorer les performances~\cite{Paleczny2001,Kotzmann2008,Haubl2011}. La conception de la JVM possède un biais initial qui
favorise les langages statiques fortement typés. L'intérêt croissant envers les langages dynamiques
sur la JVM a amené l'apparition de \texttt{invokedynamic} et de l'API de support
\texttt{java.lang.invoke}~\cite{Rose2009}. Ceci facilite l'implémentation de ces langages tout en
donnant des points d'entrée pour les optimisations adaptatives du JIT~\cite{Thalinger2010}. De
nouveaux usages de \texttt{invokedynamic} commencent à apparaître~\cite{Appeltauer2010}, et le
support des \emph{lambdas} dans Java 8 devrait s'appuyer dessus\footnote{ 
\url{http://openjdk.java.net/projects/lambda/}}.

\paragraph{Mise à jour dynamique de logiciels} La fonctionnalité de mise à jour dynamique de
logiciel n'est pas une nouvelle idée et a été longuement étudiée par de nombreux
travaux~\cite{Frieder1991,Hicks2005}. Les techniques mises en {\oe}uvre dépendent fortement
du langage de programmation ou de la plate-forme d'exécution ciblée. Des patchs dynamiques peuvent
être appliqués au niveau du système d'exploitation comme dans~\cite{Arnold2009} et~\cite{Palix2011}.
Plus proche de la machine virtuelle Java, les travaux de~\cite{Wurthinger2010} permettent de
modifier le code Java à l'exécution en supportant l'ajout ou le retrait de champs et méthodes. Cette
approche requiert toutefois une machine virtuelle modifiée. Les mêmes auteurs utilisent cette même
machine virtuelle modifiée et appliquent une programmation par aspects dynamique
dans~\cite{Wurthinger2011}. Il est important également de mentionner que la redéfinition des classes
passe par un rechargement dans la machine virtuelle. C'est également le cas dans des approches
comme Byteman~\cite{Dinn2011} qui travaille avec des définitions stables des classes, les recharge
et perd les optimisations accumulées par le JIT.  D'autres approches pour la mise à jour dynamique
utilisent des machines virtuelles modifiées. \emph{JnJVM} offre une machine virtuelle à composants
dans laquelle non seulement des aspects peuvent être injectés dynamiquement pour modifier les
applications, mais la machine virtuelle elle-même est sujette aux modifications \cite{jnjvm08}.
\emph{JVolve} est une autre machine virtuelle, dérivée de \emph{Jikes RVM}, et qui supporte
l'évolution d'applications \cite{Subramanian09}.

\section{Conclusions et perspectives}

\subsection{Conclusion}

Cet article a présenté JooFlux, un agent JVM qui permet à la fois le remplacement dynamique
d'implémentations de méthodes et l'injection d'aspects. Par rapport aux approches existantes,
JooFlux prend un chemin original en utilisant \texttt{invokedynamic}. Les jeux de tests montrent que
le sur-coût de JooFlux est négligeable pour les invocations de méthodes, et modeste pour les aspects.
Dans tous les cas, JooFlux offre de meilleures performances que des approches similaires tels des
outils d'AOP ou des langages dynamiques. De plus, JooFlux ne nécessite pas de rechargement de
définitions de classes, ce qui préserve les optimisations du JIT pour tous les sites d'appel non
concernés. Il ne nécessite pas non plus de JVM ad-hoc. La fine couche d'indirection introduite
ne demande pas de résolution d'entrée de tableau associatif ou de garde, ce qui aide grandement
l'optimisation adaptative. Enfin, JooFlux opère de façon transparente au niveau des sites d'appel,
et il ne nécessite pas de langage dédié pour spécifier les changements.

\subsection{Perspectives}

JooFlux est  pour l'heure  un prototype  de recherche  qui démontre  comment \texttt{invokedynamic}
peut être utilisé  à d'autres fin que la mise  en oeuvre d'un langage dynamique sur  la JVM. Tandis
qu'il valide  notre approche sur  des applicatifs de  taille modeste, il  reste à travailler  à son
usage sur un plus large panel d'applications pour la JVM. 

Au vu des performances prometteuses de JooFlux, notre volonté est poursuivre l'étude de ses
applications dans des contextes spécialisés tels le contrôle de ressources, les architectures
multi-tenantes ou encore les systèmes modulaires dynamiques pour l'\emph{Internet des objets}, sans
devoir passer par des abstractions ou plates-formes telles
OSGi\footnote{\url{http://www.osgi.org/}}.  Si nos expérimentations ont pour le moment été effectuée
avec \emph{Hotspot}, nous envisageons de développer les études sur d'autres machines virtuelles Java
supportant \texttt{invokedynamic}. \emph{MaxineVM} est à ce titre une cible intéressante car elle
est développée en Java~\cite{Wimmer12}. Nous pourrions ainsi étudier l'injection d'aspects et le
remplacement dynamique de code au sein même d'une machine virtuelle et de ses composants tels le
ramasse-miettes mémoire ou l'interpréteur, de façon analogue aux travaux menés dans~\cite{jnjvm08}.
Sur le plan des fonctionnalités, nous souhaitons ajouter le support des modifications multiples de
manière transactionnelle. Il serait également intéressant de pouvoir vérifier statiquement
l'applicabilité des modifications avant de les opérer. 

\subsection{Disponibilité}

JooFlux est disponible sous la forme d'un projet open source
\url{https://github.com/dynamid/jooflux}. Il est mis à disposition selon les termes de la
\emph{Mozilla Public License Version 2.0}\footnote{\url{http://www.mozilla.org/MPL/2.0/}}. La
version du prototype de JooFlux utilisée pour la rédaction de cet article correspond au \emph{tag}
Git annoté \texttt{r0}. Nous encourageons la communauté des chercheurs et praticiens à nous remonter 
tout problème, ainsi qu'à contribuer des correctifs et améliorations.

\vspace*{-0.5em}

{\small
\bibliography{biblio}
}

\end{document}